**Limitations of Classical Force Fields for Metal Coordination Modes in Proteins: A Multilevel Study of $Ca^{2+}$ in Integrin $\alpha_V\beta_3$**

Andrea Levy[1,†], Giulia Frigerio[2,3,†], Jules Grollier[1], Paulo Siani[2,3], Cristiana Di Valentin[2,3,*], Ursula Rothlisberger[1,*]

1. *Laboratory of Computational Chemistry and Biochemistry, École Polytechnique Fédérale de Lausanne, Lausanne, CH-1015, Switzerland*
2. *Department of Materials Science, University of Milano-Bicocca, Via R. Cozzi 55, 20125 Milan, Italy*
3. *BioNanoMedicine Center NANOMIB, University of Milano-Bicocca, 20125 Milan, Italy*

[†] These authors equally contributed to this work.

* Corresponding authors: cristiana.divalentin@unimib.it, ursula.roethlisberger@epfl.ch

**Abstract**

Standard biomolecular force fields often present limitations in modeling metal coordination modes. Here, we combined classical and QM/MM molecular dynamics simulations to investigate the $Ca^{2+}$-mediated binding of cRGD to integrin $\alpha_V\beta_3$. The results demonstrate the inherent limitations of fixed-point-charge force fields in reproducing asymmetric binding modes and highlight the value of QM-based multilevel approaches to assess the correctness and accuracy of FF models and capture metal coordination modes in complex biomolecular systems.

## 1) Introduction

Molecular dynamics (MD) simulations based on classical fixed-point-charge atomistic force fields (FFs) are a vital tool in the field of computational chemistry.[1–11] By describing interatomic interactions through pairwise additive potentials, these methods enable simulations of systems containing several hundred thousands of atoms, reaching microsecond time scales at a reasonable computational cost.[12–16] This has led to remarkable successes in biomolecular simulations.[17–19] However, the functional forms of standard biomolecular FFs impose limitations in the description of phenomena such as charge transfer and polarization effects,[20–23] metal coordination,[24–27] or cation-π interactions.[28–30] In particular, metal coordination modes involve a complex interplay between electrostatics, polarization, charge transfer, and directionality imposed by the electronic structure of the metal ion and its ligands, as well as the competition between different ligands. Therefore, metal coordination modes are especially challenging to model with fixed-point-charge FFs.[31]

Because of these limitations, it is often necessary to increase the level of theory for the whole or for part of the system to ensure that there are no artefacts due to the FF description. For example, it is possible to partition the system into a small region treated at the quantum mechanical (QM) level, with the remainder described at the molecular mechanics (MM) level, using fixed-point-charge FFs. In this hybrid QM/MM approach, the electronic structure of the chemically active region, such as metal coordination sites, is treated explicitly, while taking into account the effects of the surrounding environment at a less computationally demanding level. This multiscale approach is particularly relevant for biological systems containing metal ions, where an accurate description of metal coordination modes is crucial.

Notably, approximately 33% of protein structures deposited in the Protein Databank (PDB) contain metal ions.[32] Among these, integrins represent a large family of cell-surface receptors mediating crucial interactions between cells and the extracellular matrix. [33] Integrin $\alpha_V\beta_3$ is one of the most important and clinically relevant members of this family, as it is overexpressed in several cancer types and widely targeted for both active drug delivery [34] and anticancer therapy (e.g., cilengitide for glioblastoma [35]). Due to the large size of these protein systems, totaling more than 1600 amino acids, most ligand and drug design studies in the literature have relied on fixed-point-charge FFs.[36–40] However, integrin $\alpha_V\beta_3$ provides a relevant and challenging test case for fixed-point-charge FFs, as its ligand-binding site contains a divalent cation at the so-called Metal Ion–Dependent Adhesion Site (MIDAS). In addition, two other divalent cation sites are located near the binding pocket, namely the Adjacent to MIDAS (ADMIDAS) ion, more exposed to the solvent, and the Ligand-Induced Metal Binding Site (LIMBS), buried in the protein. The MIDAS ion directly mediates ligand recognition by coordinating the carboxylate group of the Asp residue within the (Arg-Gly-Asp) RGD motif, which is the amino acidic sequence selectively recognized by integrin $\alpha_V\beta_3$ both in natural and synthetic ligands.[34,41,42] Experimental X-ray structures reveal a *bridging* coordination mode (Figure 1), where one of the O atoms of the RGD Asp carboxylate is pointing towards the MIDAS ion and the other one to the integrin backbone.[43] In contrast, MD simulations performed at the molecular mechanics (MM) level with the fixed-point-charge FF CHARMM36m/CGenFF [5–7,44–46] by Frigerio et al.[47] provide a dynamic picture in which the binding mode interconverts between the *bridging* coordination mode and the *chelating* one, where both O atoms coordinate the MIDAS ion (Figure 1). These discrepancies between experimental and computational observations leave it unclear whether the *chelating* coordination mode observed in MM MD simulations reflects a realistic conformation accessible under aqueous biological conditions that is not captured in the crystalline state of the X-ray structures, or whether it represents an artefact of the fixed-point-charge FF. In this work, we present a detailed investigation of the metal coordination modes in the integrin $\alpha_V\beta_3$/ligand (cyclic RGD, c(RGDyK) – hereafter referred to as cRGD) complex, focusing on the MIDAS ion. By using a higher level of theoretical description, we aim to determine if the metal coordination modes observed in fixed-point-charge FF-based MD simulations likely correspond to physically relevant states or to FF artefacts. Starting from classical FF results, we employ QM/MM MD to assess the stability of the two metal coordination modes. Although the current study focuses on a single system treated with a specific FF, the results and conclusions highlight some inherent universal possible issues of fixed-point-charge FFs in modeling metal coordination modes in biological systems and illustrate how QM/MM can be used to validate and potentially improve the FF description.

In the following, we discuss the systems and simulation protocols in Section 2. We present the results for the integrin system in Section 3, as well as the impact of refining the CHARMM36m/CGenFF description. Finally, Section 4 starting from summarizing our conclusions, discusses more general implications for modeling metal coordination modes in biological systems using classical FFs based on fixed point charges.

**2) Computational details**

The model system is composed by the headpiece of integrin $\alpha_V\beta_3$ in complex with the cRGD ligand, which shares the cyclic structure characteristic of typical integrin $\alpha_V\beta_3$-selective ligands.[36,38]

The structure of the integrin $\alpha_V\beta_3$ extracellular segment was taken from the PDB (code 4MMX).[48] The fibronectin ligand was removed while the two co-crystallized waters in the binding pocket were kept and the eight co-crystallized metal ions were modeled as $Ca^{2+}$ ions, according to a common practice.[38,49,50] The protein structure was processed with the *Protein Preparation Wizard* in Maestro (release 2021-1).[51] The cRGD ligand was docked into the binding site with the rigid receptor docking protocol described in the SI, using the *Glide* software.[52–54] To reduce the computational cost of MD simulations, we considered only the β-propeller domain for chain α and β-A and the hybrid domains for chain β, which constitute the headpiece of integrin $\alpha_V\beta_3$ containing the ligand binding site and seven structural $Ca^{2+}$. This strategy was adopted also in other works.[55–59]

The protein/ligand complex was prepared for MM MD simulations using the *Ligand Reader & Modeler* CHARMM-GUI module to generate topology and FF parameters.[60] The same FF as in the previous work by some of us[47] was used: CHARMM36m FF[5,6,44] for the protein and CGenFF[7,45,46] for the cRGD ligand, removing the NBFIX correction for $Ca^{2+}$ cations, whose impact has been assessed in Ref. [47]. The system was solvated in an aqueous solution of NaCl (0.15M), using the rigid CHARMM-modified TIP3P model[5,61] in a cubic box with a side length of 125Å, with GROMACS tools[62] or CHARMM-GUI *Solution Builder* module[60,63], to obtain independent replicas with different initial solvent configurations. Three independent replicas were simulated with different initial setups: one 1 μs simulation starting from the system solvated with GROMACS tools, and two 500 ns simulations both starting from the system solvated with the CHARMM-GUI *Solution Builder* and differing in their initial velocity assignments. For each replica, a minimization of 5000 steps using the steepest descent algorithm was carried out, followed by a 2 ns NVT equilibration and three NPT runs. The temperature was set at 303.15K by V-Rescale thermostat,[64] with a coupling constant of 1.0 ps. The pressure was kept constant at 1 bar by a Parrinello-Rahman barostat[65] with a coupling constant of 5.0 ps, with fixed box angles and only the box lengths allowed to vary. The Particle Mesh Ewald (PME) solver[66] handled the electrostatic interactions with a real-space cutoff of 12.0Å, while van der Waals interactions were smoothly switched to zero using a force-switch function between an inner cutoff of 10Å and an outer cutoff of 12Å. Newton's equations of motion were integrated with the leap-frog algorithm.[67] All covalent bonds involving a hydrogen atom were constrained using LINCS.[62,68] MM MD simulations were performed using GROMACS.[62] QM/MM MD simulations were performed using the MiMiC framework,[69,70] which assigns different external programs to different subsystems. The QM subsystem (Figure 1) is composed of 247 atoms at the binding site of the cRGD ligand and contains the MIDAS ion, as well as the nearby LIMBS ion with their coordinating residues and 33 water molecules. The QM subsystem was handled by CP2K,[71,72] and was treated at DFT level with the BLYP functional[73,74] with density and relative cutoffs (750Ry and 90Ry, respectively) determined using standard procedures in CP2K[72], using a cubic box of length 40Å and the multipole Poisson solver available in CP2K to decouple QM periodic images with the use of density derived atomic point charges. [75] The MM subsystem was handled by GROMACS, [62,76,77] with the same parameters as in the previous MM MD simulations. We also converged the QM/MM parameters for the long-/short-range QM/MM electrostatic coupling in MiMiC according to Ref.[70], resulting in a short-range cutoff of $80a_0$ (~42Å) and a multipole order of 7 for the expansion of the QM density. The MD propagation (Born-Oppenheimer MD) was handled by the MD driver,

CPMD,[78] using a time step of $20\hbar/E_h$ (~0.5fs) and Nosé-Hoover thermostats[79,80] on the two subsystem (with reference temperature of 303.15K and coupling frequency 3000 $cm^{-1}$).

In the QM/MM thermodynamic integration (TI)[81–83] simulations, the MD driver handled the constraints, and a distance constraint has been imposed between the MIDAS ion and the OD2 oxygen of the cRGD ligand. Each TI window was simulated for around 1 ps, until the average constraint force reached convergence (Figures S4-S8). Configurations from the QM/MM TI windows were used as starting point for the MM TI, in which similar constraints were imposed with GROMACS. In this case, each window was simulated for longer time scales (around 1 ns) until convergence of the average constraint force.

## 3) Results and Discussion

### 3.1) MM MD simulations

Our classical MD simulations of cRGD in complex with integrin $\alpha_V\beta_3$, including a 1 μs simulation (Figure S2) and two 500 ns replicas (Figure S3) as detailed in Section 2, consistently reveal two distinct coordination modes of the MIDAS ion with the cRGD Asp carboxylate: *bridging* and *chelating* (Figure 1), which were also found in a previous work by some of us.[47] In particular, in the *bridging* mode, one carboxylate oxygen – hereafter referred to as OD1 – coordinates the MIDAS $Ca^{2+}$ ion in a monodentate fashion (distance MIDAS–OD1 around 2.0-2.5 Å), while the second oxygen – hereafter referred to as OD2 – forms a hydrogen bond with the backbone amide of Tyr122 in the β chain (distance MIDAS–OD2 around 4.0 Å). In the *chelating* mode on the other hand, both carboxylate oxygens directly coordinate the metal ion (both distances MIDAS–OD1 and MIDAS–OD2 around 2.0-2.5 Å). The average MIDAS–OD1/OD2 distances during the 1 μs classical MD are reported in Table 1.

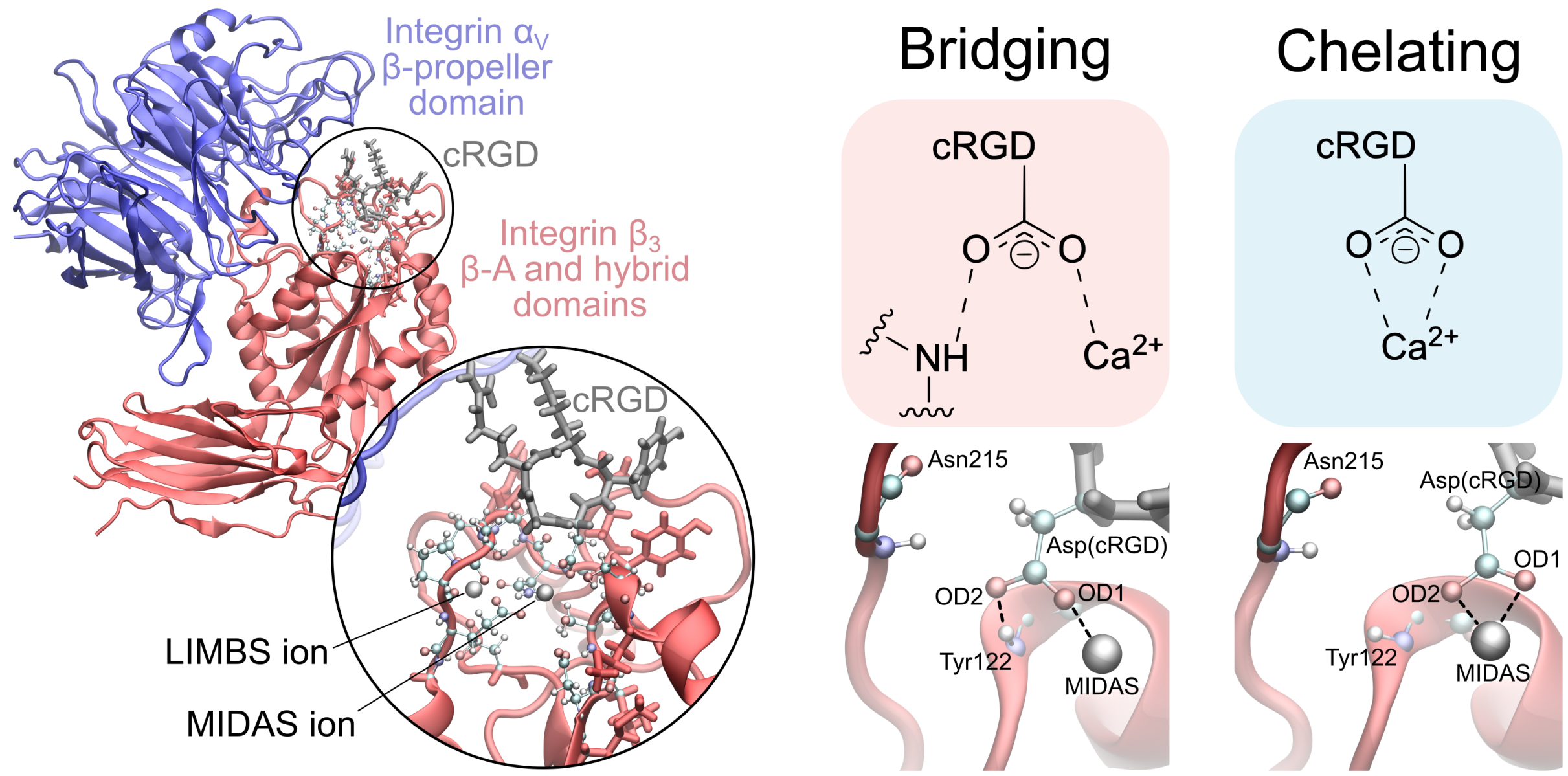

**Figure 1.** Model system used for MM and QM/MM MD simulations (left). The QM region (indicated with the circles) highlights the atoms treated at QM level in ball and stick representation (water molecules are omitted for clarity). Metal coordination modes for MIDAS ion (right). For both modes, a schematic representation is reported, as well as the initial configuration for the unconstrained QM/MM MD, where only the relevant residues are represented.

### 3.2) QM/MM MD simulations

In order to perform QM/MM MD simulations, we extracted two representative well-equilibrated configurations from the MM MD simulations, where the system is in the *bridging* (at 500 ns of the 1 μs replica) or in the *chelating* binding mode (at the end of the 1 μs replica). For both cases, we performed a typical QM/MM equilibration to relax the system at the QM/MM level: after a simulated annealing phase to reduce the energy of the QM region, dynamically, letting the MM region adapt, the system was slowly heated to reach the desired temperature of 303.15K, to match the one used for the classical MD.

After this equilibration, NVT QM/MM MD simulations have been performed for ~10ps each, and both coordination binding modes were maintained through the MD, with the average

MIDAS–OD1/OD2 distances reported in Table 1. The evolution of these distances during the QM/MM MD is reported in the SI (Figures S9 and S10).

**Table 1.** Distances during the unconstrained MM (1 μs replica) and QM/MM MD between OD1 and OD2 oxygen atoms and the MIDAS $Ca^{2+}$ ion. For each value, the standard deviation is also reported.

| Simulation | Coordination mode | MIDAS–OD1 distance (Å) | MIDAS–OD2 distance (Å) |
|---|---|---|---|
| MM MD | Bridging | 2.10 ± 0.04 | 4.0 ± 0.1 |
| | Chelating | 2.2 ± 0.1 | 2.3 ± 0.3 |
| QM/MM MD | Bridging | 2.3 ± 0.1 | 4.2 ± 0.1 |
| | Chelating | 2.5 ± 0.1 | 2.6 ± 0.2 |

Starting from the *chelating* binding mode, intermediate configurations have been generated by systematically varying distance constraints, in order to increase the OD2–MIDAS distance and to obtain starting configurations along the reaction path from the *chelating* to the *bridging* mode. Such configurations were used to perform TI,[81–83] where constrained QM/MM MD simulations are performed along the reaction path (collective variable, CV=MIDAS–OD2 distance). By integrating the average constraint force along the CV, we obtained the free energy profile reported in Figure 2, where results from the unconstrained QM/MM MD simulations have been used to add two "TI windows" with zero constraint force, in which the system is stable without the need of imposing any additional constraints.

From the results of QM/MM simulations, the two different coordinating modes appear to be both stable minima for cRGD, but among them the *bridging* binding mode, which is also the one observed in experimentally determined crystallographic structures,[43] has a free energy ~2.6 kcal/mol lower. Moreover, the two coordination minima are separated by a free energy barrier of ~5.2 kcal/mol (*bridging → chelating*), and ~2.6 kcal/mol (*chelating → bridging*). These barriers correspond to crossing times of ~1ns and ~10 ps, respectively (estimated through the Eyring–Polanyi equation[84] assuming zero re-crossings and a constant prefactor).

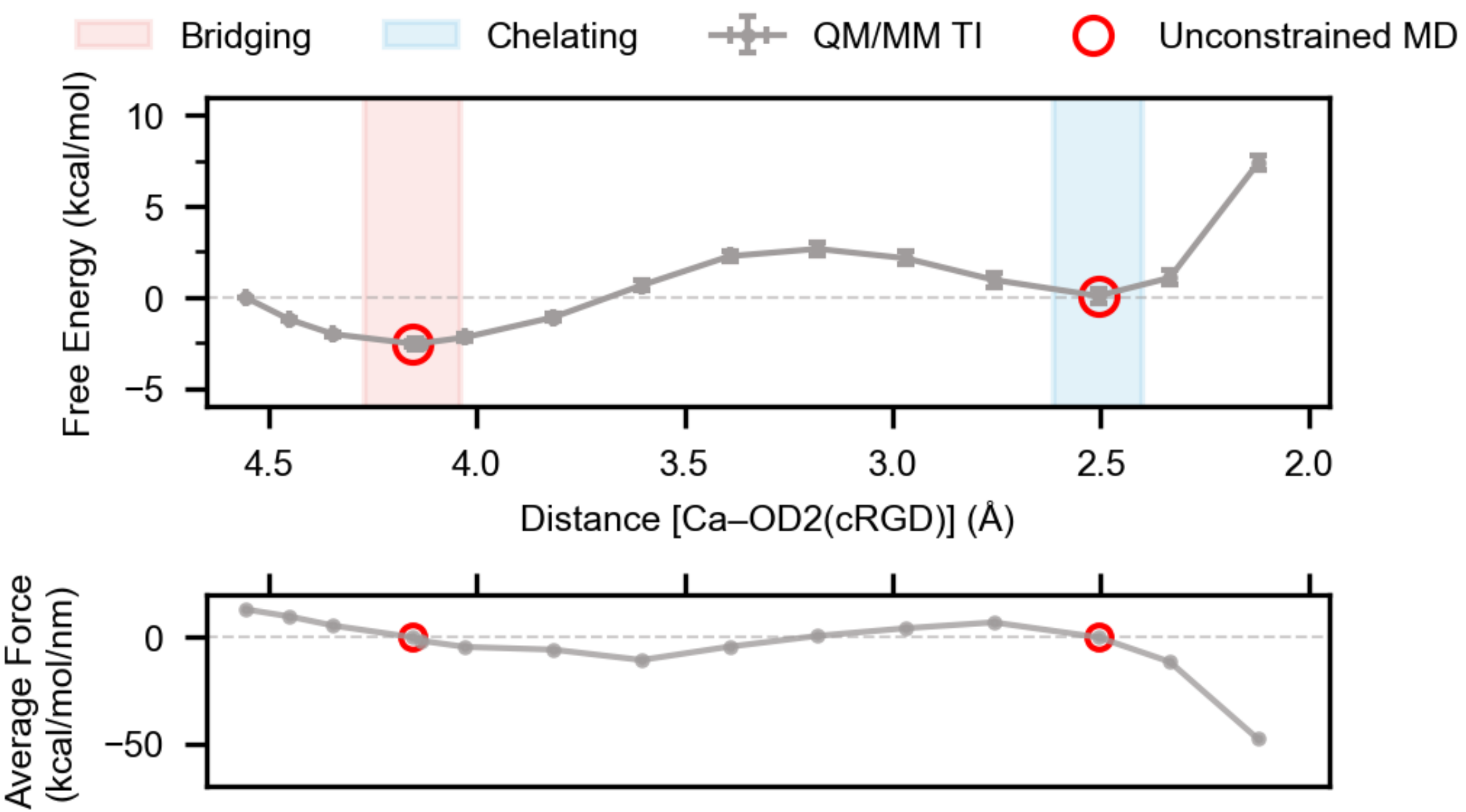

**Figure 2.** Free energy profile from QM/MM TI (upper panel) with corresponding average constraint forces (lower panel). Unconstrained QM/MM MD simulations have been used to add zero-force points at the average unconstrained distance. For each unconstrained simulation, the standard deviation of the distance is also indicated in the plot as a colored area.

### 3.3) Atomic charge analysis

During the QM/MM MD simulations, we made use of the xDRESP method, recently introduced within the MiMiC framework,[85] in which the electrostatic potential generated from the QM subsystem on the surrounding MM atoms is used for determining atom-centered electric multipoles. The results of this analysis can be used to have a qualitative understanding of how the electron density is affected during a reaction. Notably, in our case, the atomic charge of the MIDAS ion remains stable during the unconstrained MD in both the *bridging* (0.47±0.03 *e*) and the *chelating* (0.48±0.03 *e*) conformation (Figure 3). In both cases, its charge is lower than the formal oxidation state of $Ca^{2+}$, which is a general phenomenon for different charge schemes because electron-donating ligands, such as the carboxylic group in this case, transfer part of their negative charge to the metal ion. The xDRESP charges can be even lower than those obtained with other charge schemes, as they use Hirshfeld charges[86] as restraints during the fit which are based on the isolated-atom densities (neutral). Nevertheless, the xDRESP point charges of the two oxygen atoms of the cRGD ligand, involved in the *chelating* or *bridging* mode, present larger differences in the two metal coordination modes, especially in the case of the OD1 oxygen, which is the one bound to the cation in both configurations and whose charge changes from -0.42±0.05 *e* in the *bridging* mode to -0.28±0.04 *e* in the *chelating* one (Figure 3).

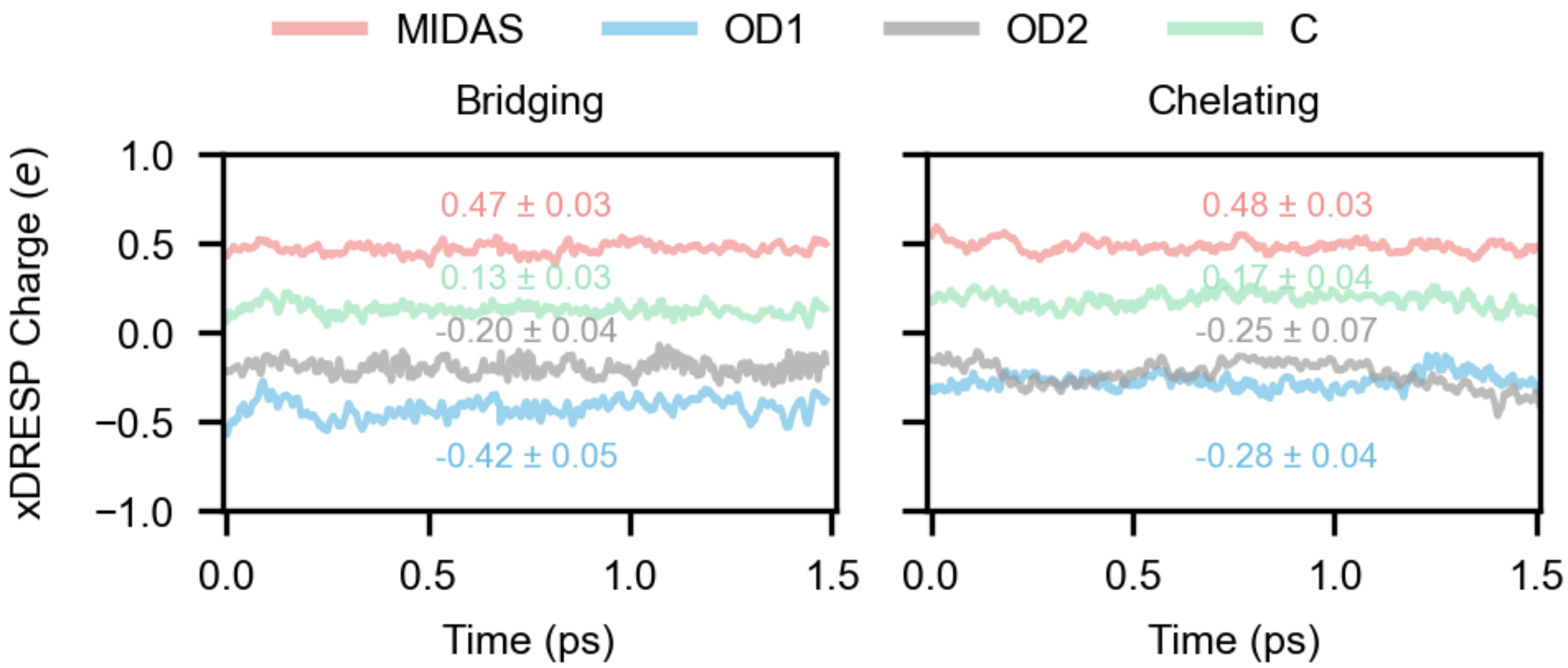

**Figure 3.** xDRESP charges from the last ~1.5ps of the unconstrained QM/MM MD for the two coordination modes. For both cases, the xDRESP charge of the MIDAS ion (red), OD1 oxygen (blue), OD2 oxygen (gray) and carboxylate carbon (green) are reported. The corresponding average values and standard deviations are also reported, using the same color code.

In comparison, the FF used during the MM MD simulations assigns the same charge (–0.776 *e*) for the two oxygens, as they are chemically equivalent in the isolated cRGD ligand (Table 2). To further investigate the dependence of the point charges from the coordination mode, we performed RESP fitting on two model systems designed to mimic the chemical environment of the two coordination modes (Figure S11), where we constrained all the point charges – except the ones of the two oxygen atoms – to their value in the original FF. In particular, as the MM model uses the CGenFF,[45,46] in which initial guess partial atomic charges are estimated from MP2/6–31G(d) Merz–Kollman charges,[87,88] we performed geometry optimizations of the model systems using the same level of theory (using Gaussian16[89]), and Multiwfn for the RESP fit.[90,91]

The results from the RESP fit are reported in Table 2, where we changed the reference configurations used during the fit: *chelating vs bridging* and we also performed multi-conformation fit, where we used equal weights for both, or a 0.25–0.75 ratio, favoring the *chelating* configuration. These results represent only a limited subset of all possible charge fitting strategies, and could be extended to, e.g., allow variations of more atoms, or varying the restraint parameters for the RESP fit. However, they clearly show that large variations are present between the two conformations, and that, similarly to what was observed with the xDRESP analysis, the OD1 oxygen, which binds the MIDAS ion in both configurations, consistently presents a larger negative charge in the *bridging* coordination mode and is always more negative than the OD2 oxygen. Notably, in the xDRESP analysis, the OD1 oxygen is the only atom with such a large difference in the average xDRESP charge between the two modes among all the atoms coordinating the MIDAS ion, while the xDRESP charge of the OD2 oxygen remains similar, although its environment changes more drastically (between coordinating and noncoordinating).

**Table 2.** Effective atomic charges for the OD1 and OD2 oxygen atoms. In case of RESP fitting, the results have been obtained from model systems, imposing all the remaining charges to be the same as in the original FF.

| | Atomic charge (*e*) | |
|---|---|---|
| | OD1 | OD2 |
| Original FF (CGenFF) | -0.776 | -0.776 |
| xDRESP (*bridging* mode) | -0.42 ± 0.05 | -0.20 ± 0.04 |
| xDRESP (*chelating* mode) | -0.28 ± 0.04 | -0.25 ± 0.07 |
| RESP<br>(100% *bridging*) | -1.147 | -0.405 |
| RESP<br>(50% *bridging*, 50% *chelating*) | -1.063 | -0.489 |
| RESP<br>(25% *bridging*, 75% *chelating*) | -0.972 | -0.580 |
| RESP<br>(100% *chelating*) | -0.844 | -0.708 |

These observations suggest that, even if a fixed-point-charge model is potentially sufficient to describe the MIDAS ions, as its xDRESP charge is unchanged in the two configurations, the charges of its coordinating atoms are strongly influenced by the coordination mode. This may pose significant challenges for fixed-point-charge FFs in accurately describing these interactions, especially when the coordination mode switches dynamically.

**3.4) MM TI**

We performed additional TI at the MM level, comparing the results of the original FF with the ones using adapted charges generated from the RESP fitting. By performing MM TI on similar windows as the QM/MM TI (Figure 4), we concluded that the two configurations are indeed minima also for the original FF, separated by a similar barrier (~5kcal/mol) as the one obtained from QM/MM, but the original FF favors the *chelating* mode by ~1.6kcal/mol. Interestingly, small changes in the charge of the two oxygen atoms can drastically change the results from the FF simulations: for instance, when using charges fitted from the *bridging* mode only, the *chelating* mode ceases to be a free energy minimum. This is the case also when using charges from the multi-configuration fit with equal weights for the two configurations. However, when the ratio of the two is changed, assigning a larger weight to the *chelating* coordinating mode, both configurations turn out to be stable minima. Moreover, it is possible to observe a profile that is qualitatively more similar to the QM/MM TI: the two minima are still separated by a free energy barrier of ~5kcal/mol (*bridging* → *chelating*), and the *bridging* mode is energetically favored (by ~5kcal/mol). Nevertheless, the back barrier (*chelating* → *bridging*) is incorrect, and the free energy difference between the two modes is twice the QM/MM one. Additionally, the location of the minima is incorrect, especially for the *bridging* mode which is ~0.5Å lower for the FF-based TI profile.

As mentioned in the previous section, the considered modifications of the FF charges represent only a limited subset of possible charge fitting strategies and could be extended further. Nevertheless, the results of this section show how sensitive the metal coordination properties of a system can be to these variations, with the possibility for a conformation (e.g., the *chelating* one in this case) to become even unstable.

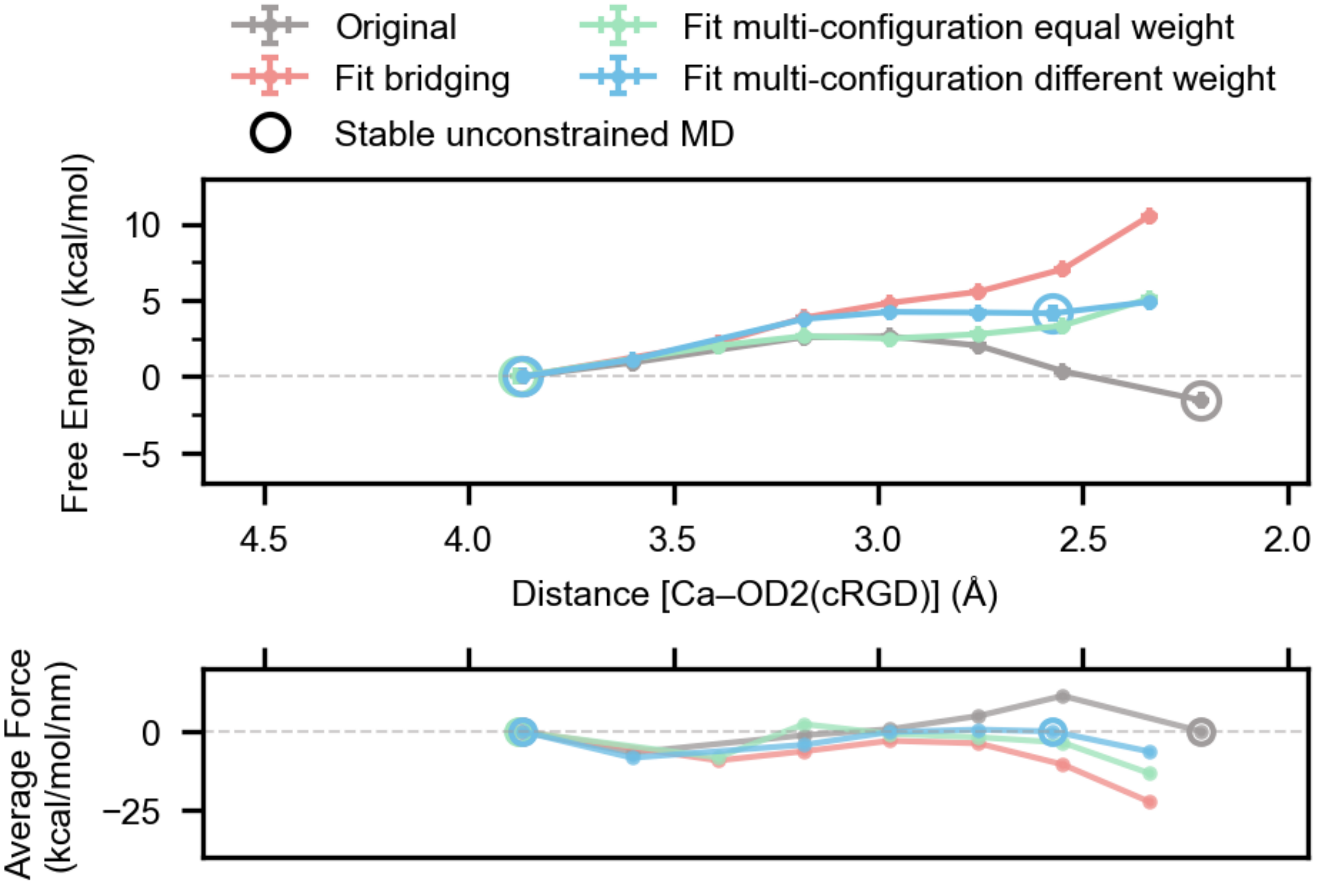

**Figure 4.** Free energy profile from MM TI (upper panel) and corresponding average constraint forces (lower panel) for different variations of the charge of the OD1 and OD2 oxygen atoms (using charges from Table 2). Unconstrained MM MD simulations have been used to add zero-force points for the stable conformations.

## 4) Conclusions and Outlook

Metal coordination modes arise from a delicate balance of electrostatics, polarization, charge redistribution, and geometric directionality. Accurately capturing this balance remains challenging for classical FFs based on fixed point charges, particularly in a dynamic environment where multiple coordination geometries may be accessible. In this work, we address this problem in the case of $Ca^{2+}$-mediated cRGD binding to the integrin $\alpha_V\beta_3$ protein. By combining classical and QM/MM MD simulations, we aim to assess whether the metal coordination modes observed in FF-based simulations correspond to physically relevant states or arise from artefacts of the FF description.

For this system, classical MD simulations predicted two distinct coordination modes, namely *chelating* and *bridging*, among which only the latter is supported by available experimental data.[43] QM/MM MD simulations show that both modes correspond to free energy minima with the experimentally observed *bridging* mode being thermodynamically favored by ~2.6kcal/mol, and that the two are separated by a free energy barrier of ~5.2kcal/mol to go from the *bridging* to the *chelating* mode, and ~2.6kcal/mol for the opposite direction. Moreover, charge analyses of the QM/MM MD runs based on the xDRESP method as well as RESP fitting on model systems, reveal that although the effective charge of the $Ca^{2+}$ ion remains essentially unchanged, the electronic distribution on the O atoms of the coordinating ligand is strongly affected by the coordination geometry. Notably, small variations in the MM charges of the two coordinating oxygen atoms of the cRGD ligand are sufficient to significantly alter the free energy landscape. Moreover, it should be noted that, besides electrostatics, the short-range repulsive component of the van der Waals interaction, arising from Pauli repulsion, may also affect the free energy profile, especially the location of the equilibrium distances. In the FF, this is primarily governed by the Lennard-Jones σ or $R_{min}$ parameter, and changing those values could improve the agreement with QM/MM results. Nevertheless, in this work these parameters were intentionally left unchanged to isolate the effect of the electrostatics. Together, our results demonstrate that the stability of metal coordination modes is most sensitive to electronic effects localized on the coordinating atoms, rather than on the metal ion itself, in agreement with what was previously highlighted for $Mg^{2+}$ by Casalino et al.[31], suggesting that the common practice of developing site-specific ion parameters does not reflect the physical origin of these effects.

More broadly, our findings highlight intrinsic limitations of classical FFs based on fixed point charges in modeling metal coordination modes in biological systems especially when asymmetric binding modes are relevant. QM/MM MD offers a potential way to assess and validate FF predictions, but its computational cost limits the accessible time scales compared to classical MD. Hence, QM/MM alone cannot fully address all challenges, such as the need for long simulation times. Alternative strategies include fluctuating charge models[92–94] and polarizable FFs,[69–72] in which the charges of the atoms coordinating the metal can adapt to the coordination environment. To limit the computational effort, these can be performed within multilevel schemes analogous to QM/MM, i.e., where only a limited region of the system is treated with higher accuracy (e.g. with a polarizable FF). In this context, QM/MM results can be used as a reference for the parametrization of polarizable FFs for the specific metal coordination modes of interest, e.g., via a force matching approach[95–97], or for machine learning (ML) potentials.[98–100] In practice, the application of such multiscale descriptions requires flexible simulation frameworks. Platforms such as MiMiC, designed to support different multilevel descriptions, can facilitate the transitions between QM/MM and, e.g., a polarizable MM/fixed-point-charge MM combined representation or ML/MM, providing an important foundation for improving the treatment of metal coordination modes in large biomolecular systems.

## Supporting Information

Details on the system preparation via docking, as well as additional figures not included in the main text from MM and QM/MM molecular dynamics simulations.

## Data availability

The data generated for this work are publicly available on Zenodo under https://doi.org/10.5281/zenodo.20844406, including Jupyter Notebooks to reproduce the analysis and generate the plots presented in the manuscript.

## Acknowledgements

This work has been supported by the Swiss National Science Foundation (grants 200020–185092 and 200020–219440) and used computing time from the Swiss National Computing Centre CSCS. The research leading to these results has received funding also from the European Union – NextGenerationEU through the Italian Ministry of University and Research under PNRR – M4C2-I1.3 Project PE_00000019 "HEAL ITALIA" to Prof. Cristiana Di Valentin CUP H43C22000830006 of the University of Milano Bicocca.

**Supporting Information**

# Limitations of Classical Force Fields for Metal Coordination Modes in Proteins: A Multilevel Study of $Ca^{2+}$ in Integrin $\alpha_V\beta_3$

Andrea Levy[1,†], Giulia Frigerio[2,3,†], Jules Grollier[1], Paulo Siani[2,3], Cristiana Di Valentin[2,3,*], Ursula Rothlisberger[1,*]

1. *Laboratory of Computational Chemistry and Biochemistry, École Polytechnique Fédérale de Lausanne, Lausanne, CH-1015, Switzerland*
2. *Department of Materials Science, University of Milano-Bicocca, Via R. Cozzi 55, 20125 Milan, Italy*
3. *BioNanoMedicine Center NANOMIB, University of Milano-Bicocca, 20125 Milan, Italy*

† These authors equally contributed to this work.

* Corresponding authors: cristiana.divalentin@unimib.it, ursula.roethlisberger@epfl.ch

**Ligand Docking**

The integrin $\alpha_V\beta_3$ was processed using the *Protein Preparation Wizard*,[51] while the ligand was prepared using the *LigPrep* tool and the grid generated using the *Receptor Grid Generation* tool. The rigid-receptor docking calculations were carried out at extra precision (XP) using the *Glide* software.[52–54] We obtained 6 docking poses for the cRGD in the binding pocket of integrin $\alpha_V\beta_3$, which were very close in docking score. To choose one, we used two quantitative criteria taking as reference the RGD sequence of cilengitide, a cyclic peptide that was co-crystallized with the integrin $\alpha_V\beta_3$ (PDB:1L5G).[43] The first criterion involves the values of the three geometrical criteria proposed by Kostidis et al.[101] and by Othman et al.[102] to assess the biological activity of RGD-based ligands. These three criteria are named $d_1$, $d_2$, and $\theta$. $d_1$ is the distance between $C_\zeta$ and $C_\gamma$ of Arg and Asp, respectively. $d_2$ is the distance between the $C_\beta$ of both Arg and Asp. $\theta$ is the angle between $C_\beta$, $C_\alpha$ and $C_\beta$ of Arg, Gly and Asp, respectively. To measure the deviation of the docking poses with respect to the values measured for the co-crystallized cilengitide, the following normalized Euclidian distance, named *Score*, was defined

$$Score = \sqrt{\left(\frac{d_1 - d_{1,0}}{d_{1,0}}\right)^2 + \left(\frac{d_2 - d_{2,0}}{d_{2,0}}\right)^2 + \left(\frac{\theta - \theta_0}{\theta_0}\right)^2}$$

where $d_{1,0}$, $d_{2,0}$ and $\theta_0$ are values for the co-crystallized cilengitide.

The second criterion compares the structure of the RGD motif of each docking pose with respect to the cilengitide one using the root mean square deviation (RMSD) computed over the heavy atoms of this common RGD motif, taking as a reference the atoms of this motif in the cilengitide crystal structure.

The RMSD and the *Score* were normalized using the standard formula:

$$X_{norm} = \frac{X - X_{\min}}{X_{max} - X_{\min}}$$

where $X_{norm}$ is the normalized value of X, X is the original value, $X_{min}$ is the minimum value in the dataset, and $X_{max}$ is the maximum value in the dataset. X represents either the RMSD or *Score* calculated for each docking pose normalized to a range between 0 and 1.

Based on the normalized RSMD and *Score* shown in Figure S1, pose 6 was selected and used as the starting point structure in MM MD simulations.

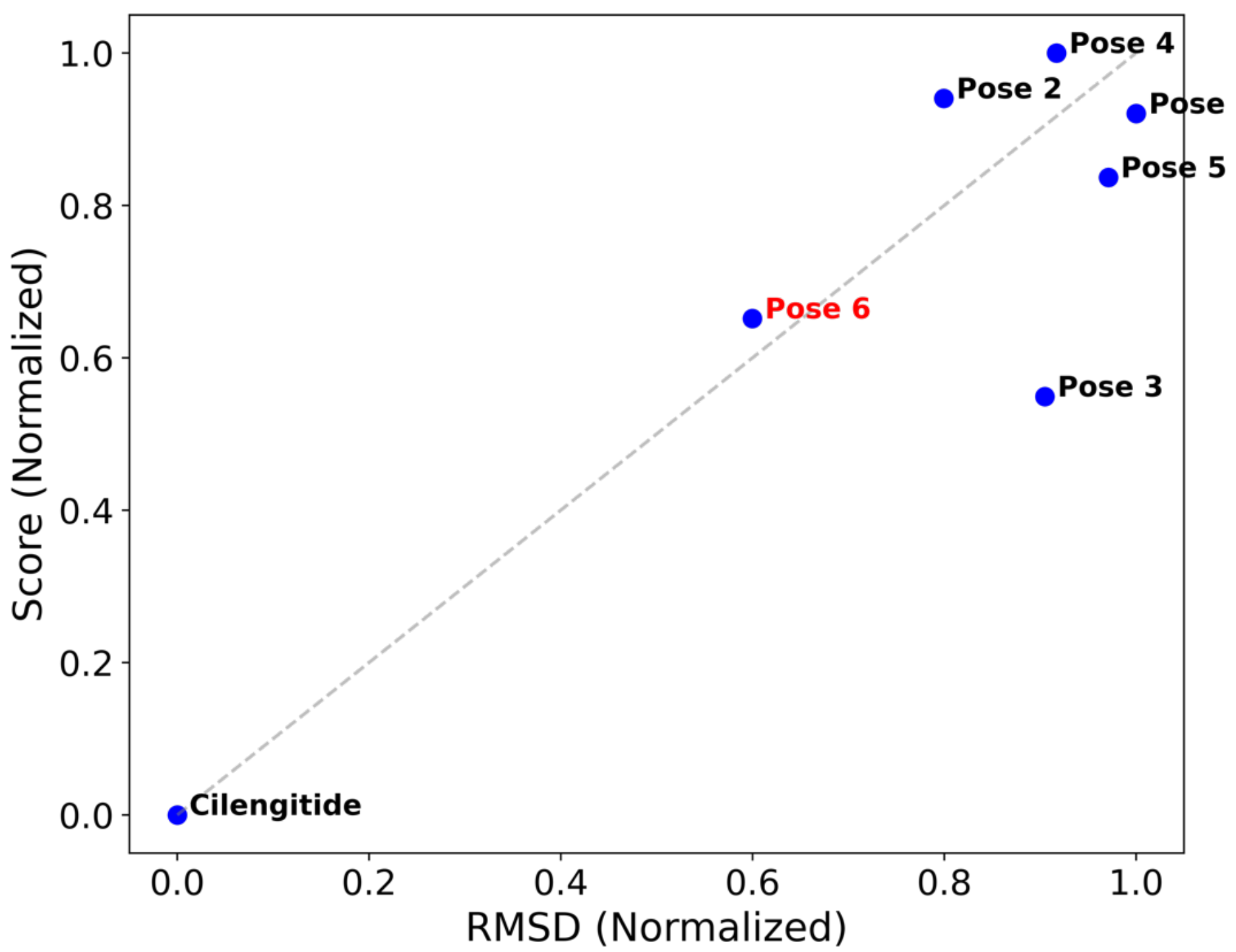

**Figure S1.** *Score* of the docking poses against the RMSD of the docking poses, where co-crystallized cilengitide is the reference structure.

## Additional figures

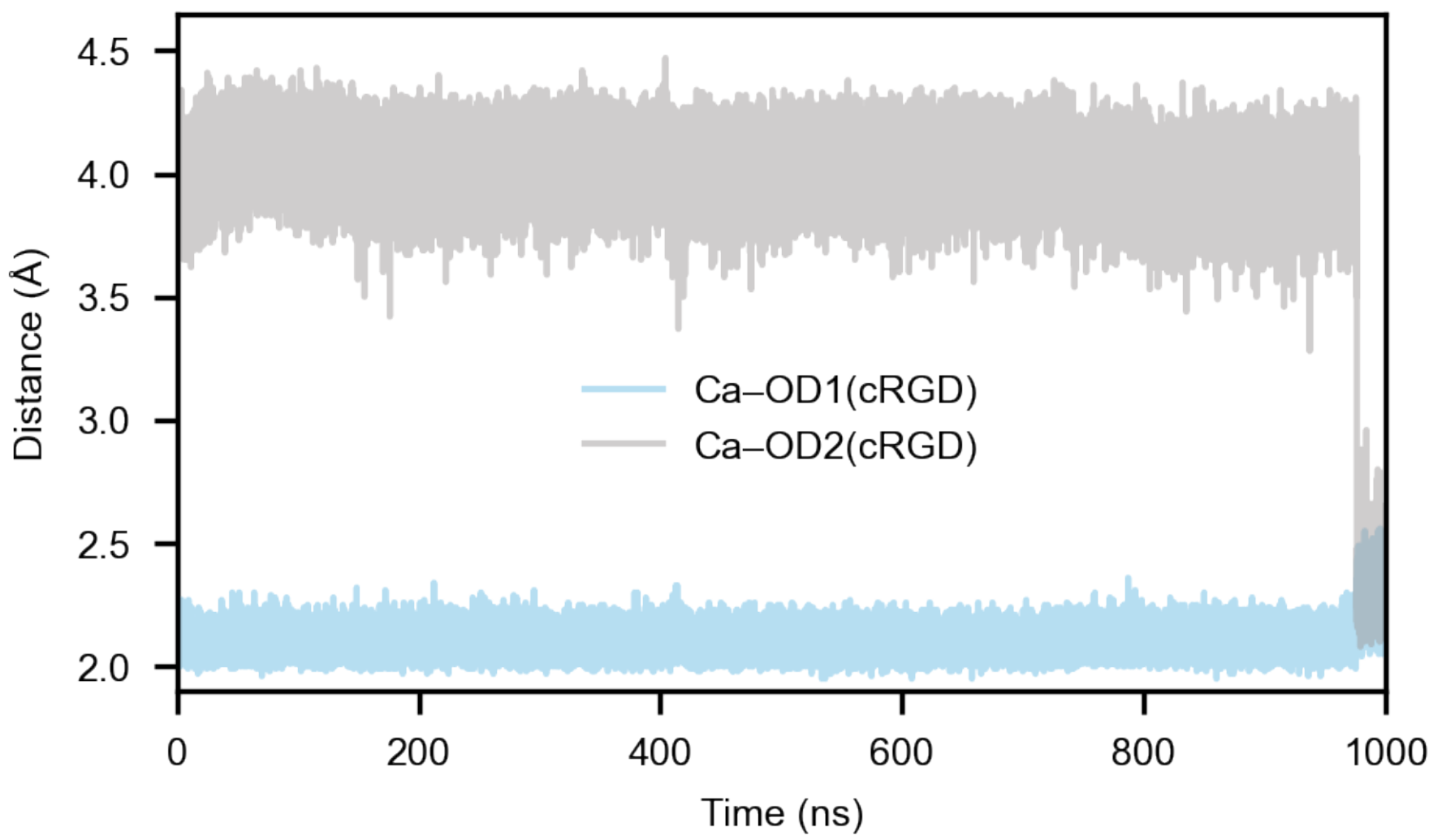

**Figure S2.** MIDAS–OD1 and MIDAS–OD2 distances during the 1µs MM MD.

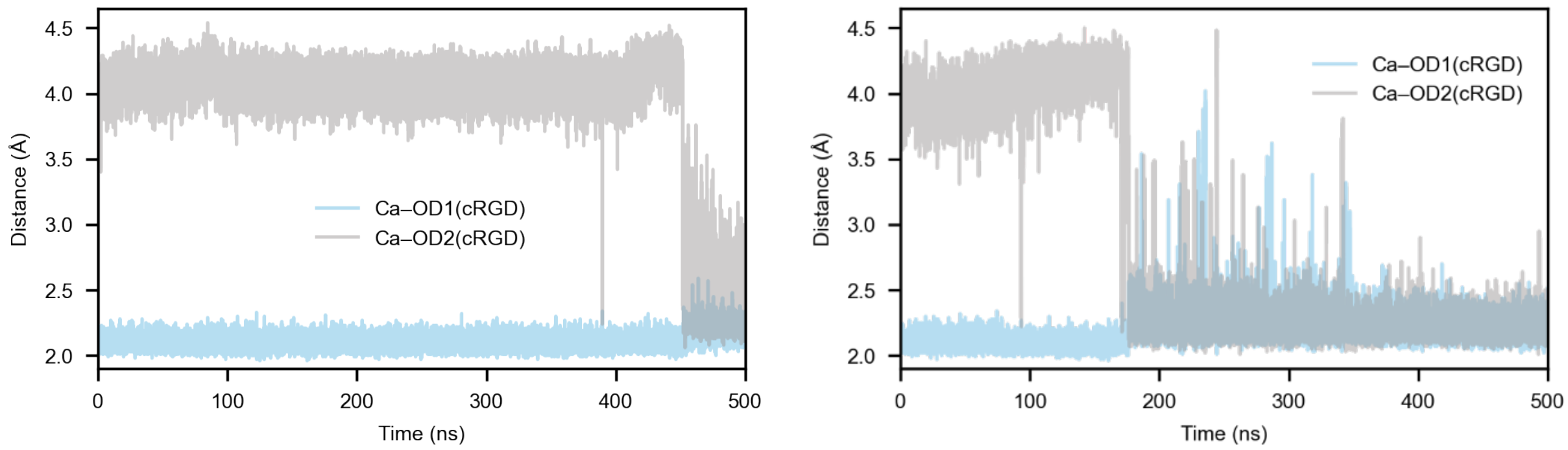

**Figure S3.** MIDAS–OD1 and MIDAS–OD2 distances during two 500ns replicas of the MM MD.

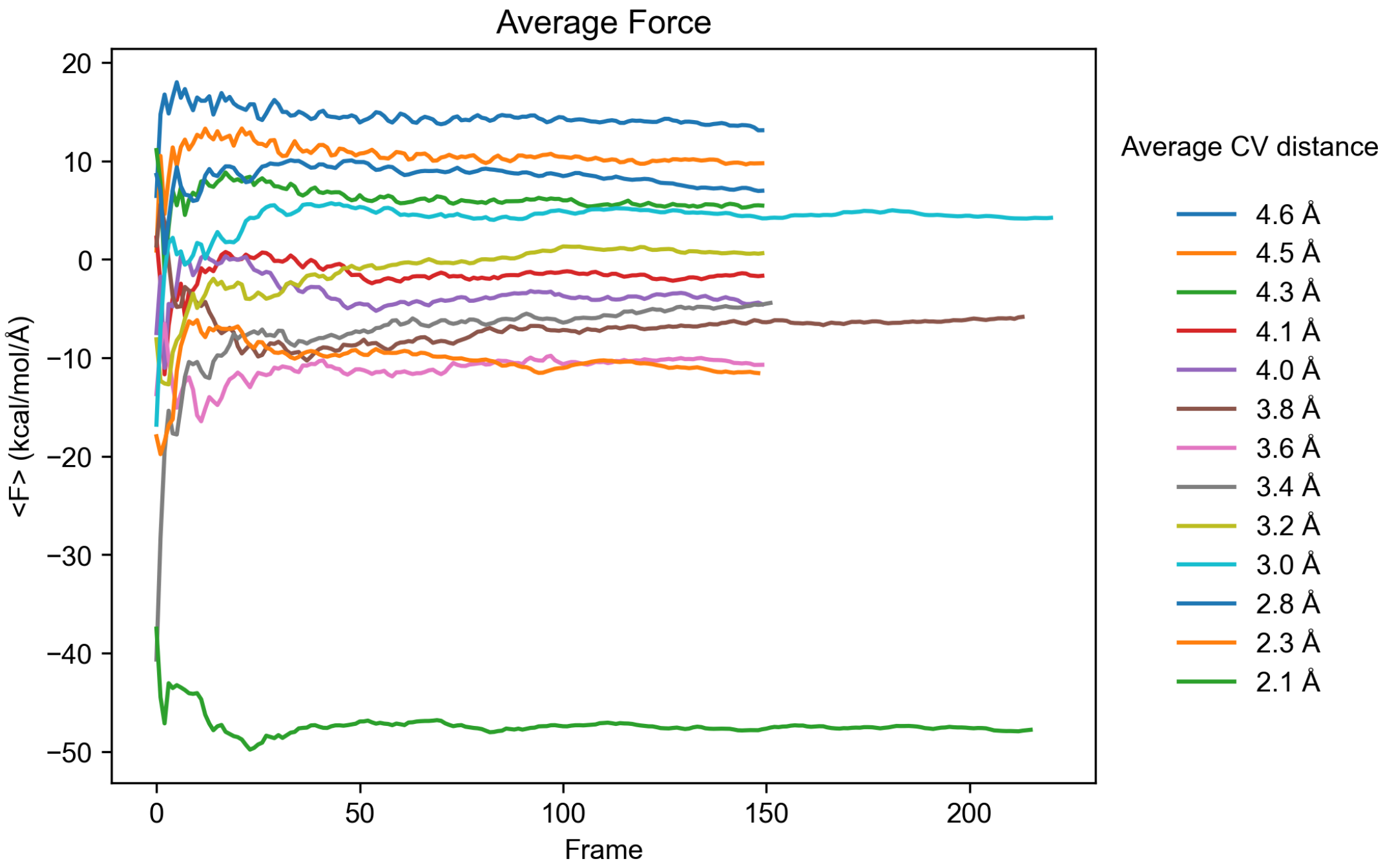

**Figure S4.** Convergence of the average constraint force for the QM/MM TI windows. Constraints are saved every 10 steps, i.e., every ~5fs.

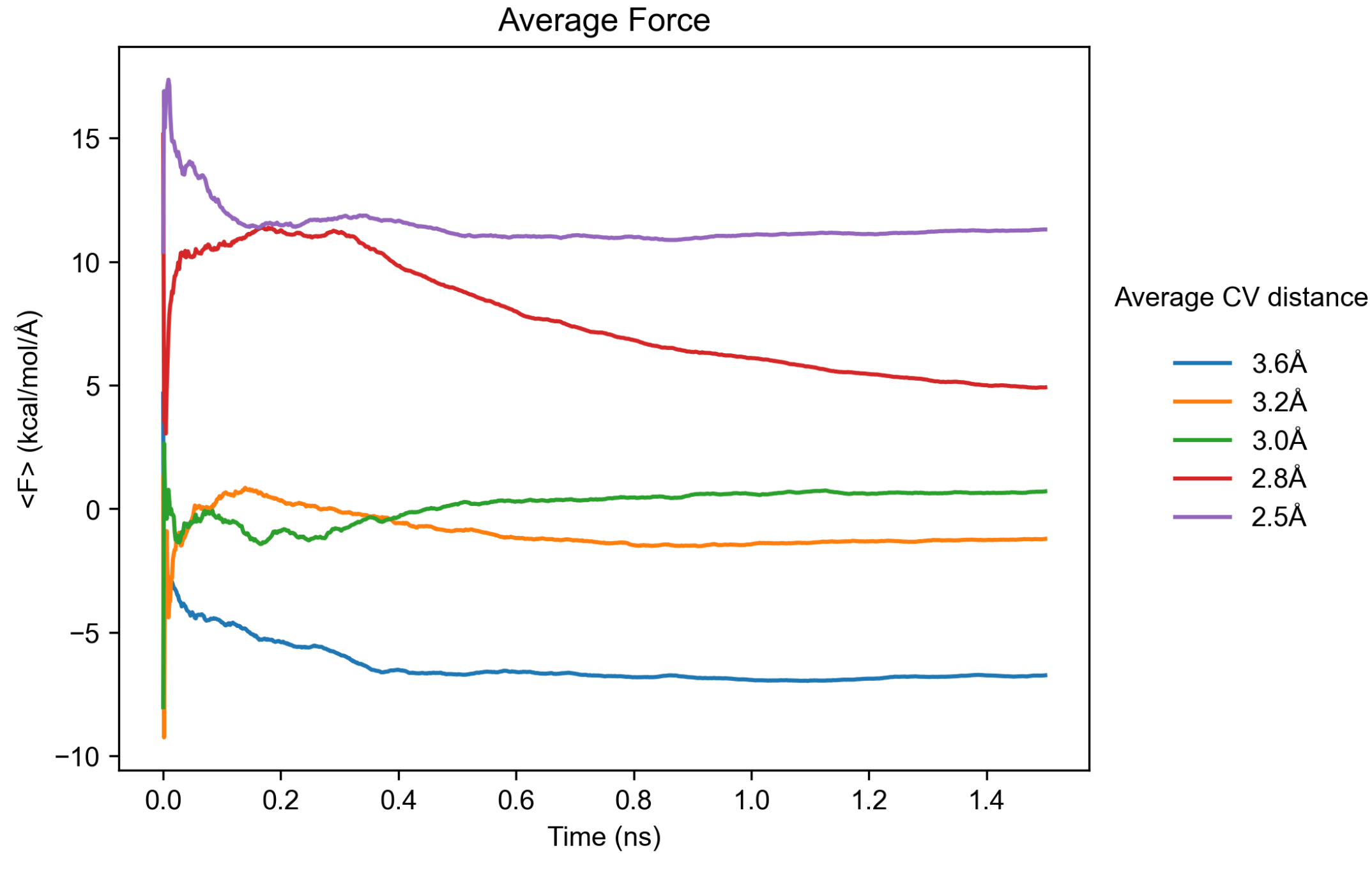

**Figure S5.** Convergence of the average constraint force for the MM TI windows with the original FF. Constraints are saved every 50fs.

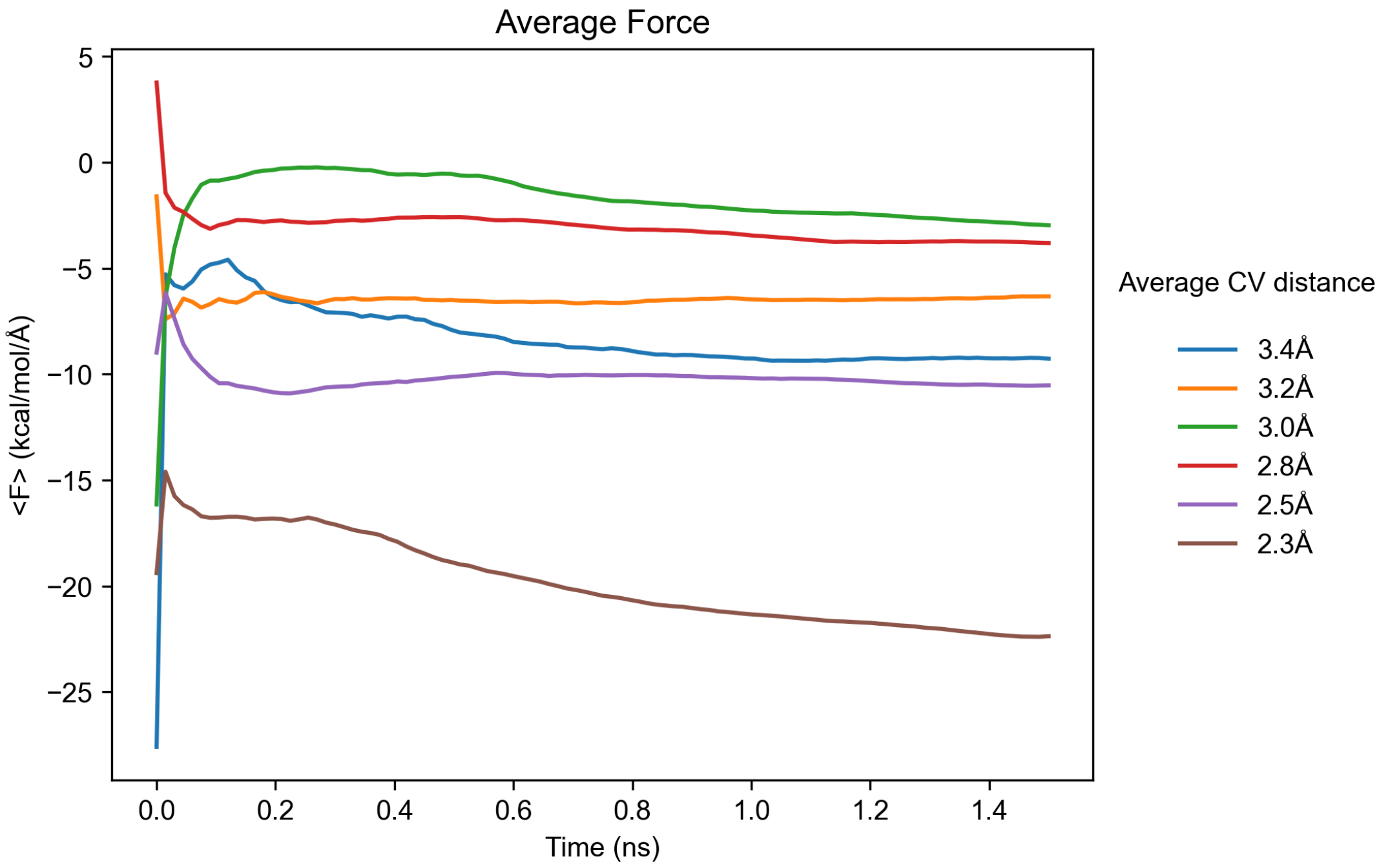

**Figure S6.** Convergence of the average constraint force for the MM TI windows with the modified FF (fitted to the *bridging* conformation). Constraints are saved every 50fs.

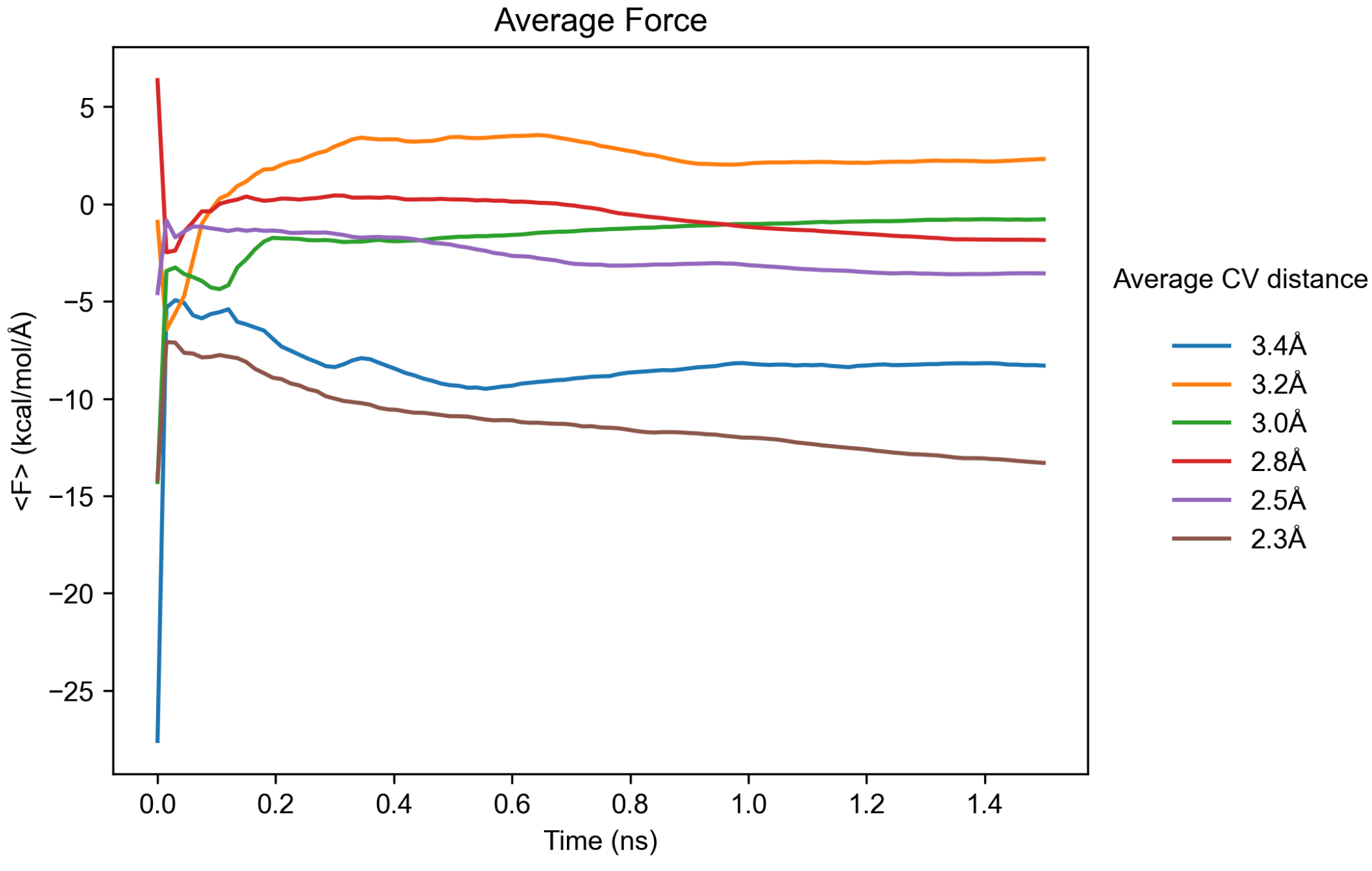

**Figure S7.** Convergence of the average constraint force for the MM TI windows with the modified FF (multiconfiguration fit – 50% *bridging*, 50% *chelating*). Constraints are saved every 50fs.

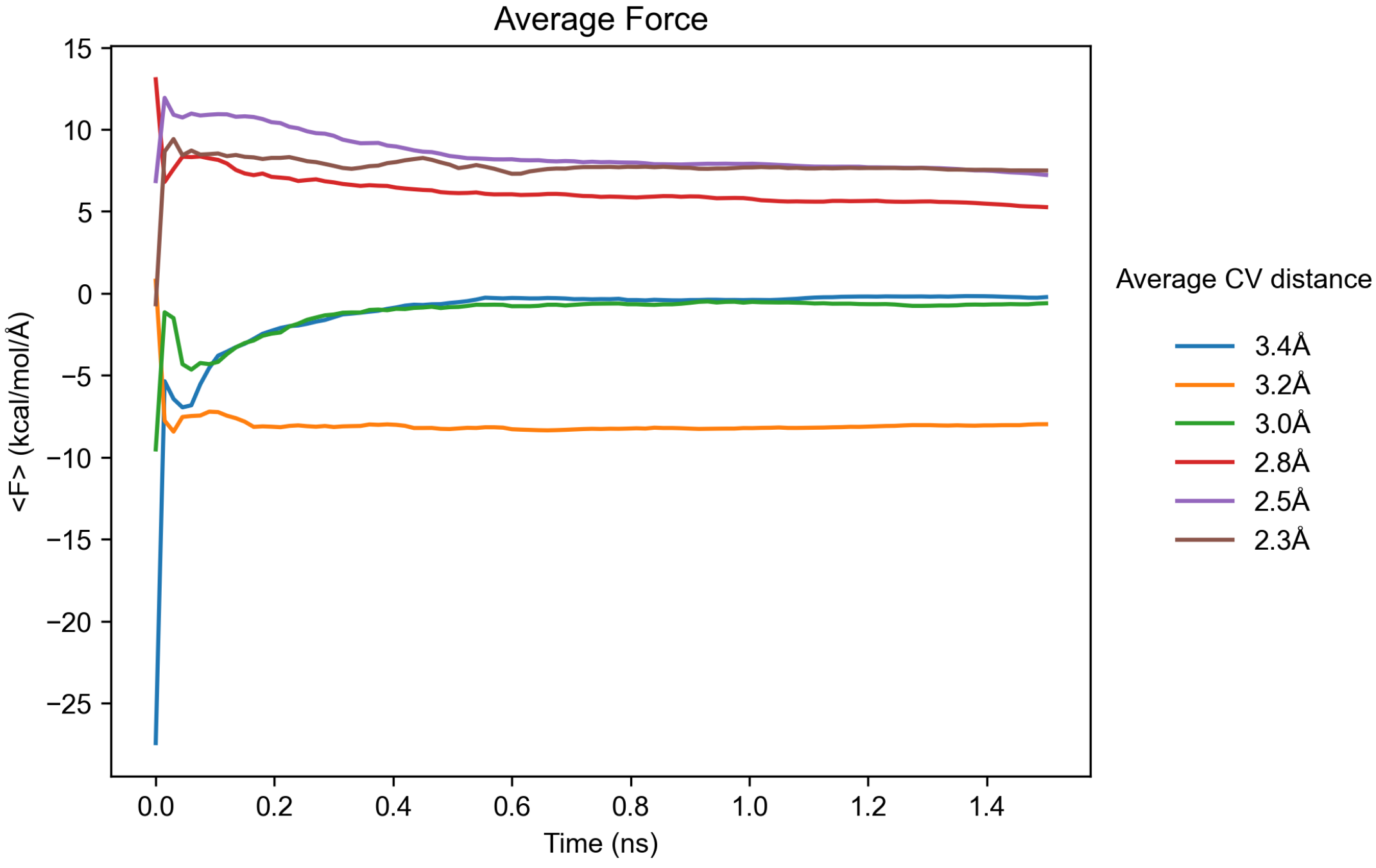

**Figure S8.** Convergence of the average constraint force for the MM TI windows with the modified FF (multiconfiguration fit – 25% *bridging*, 75% *chelating*). Constraints are saved every 50fs.

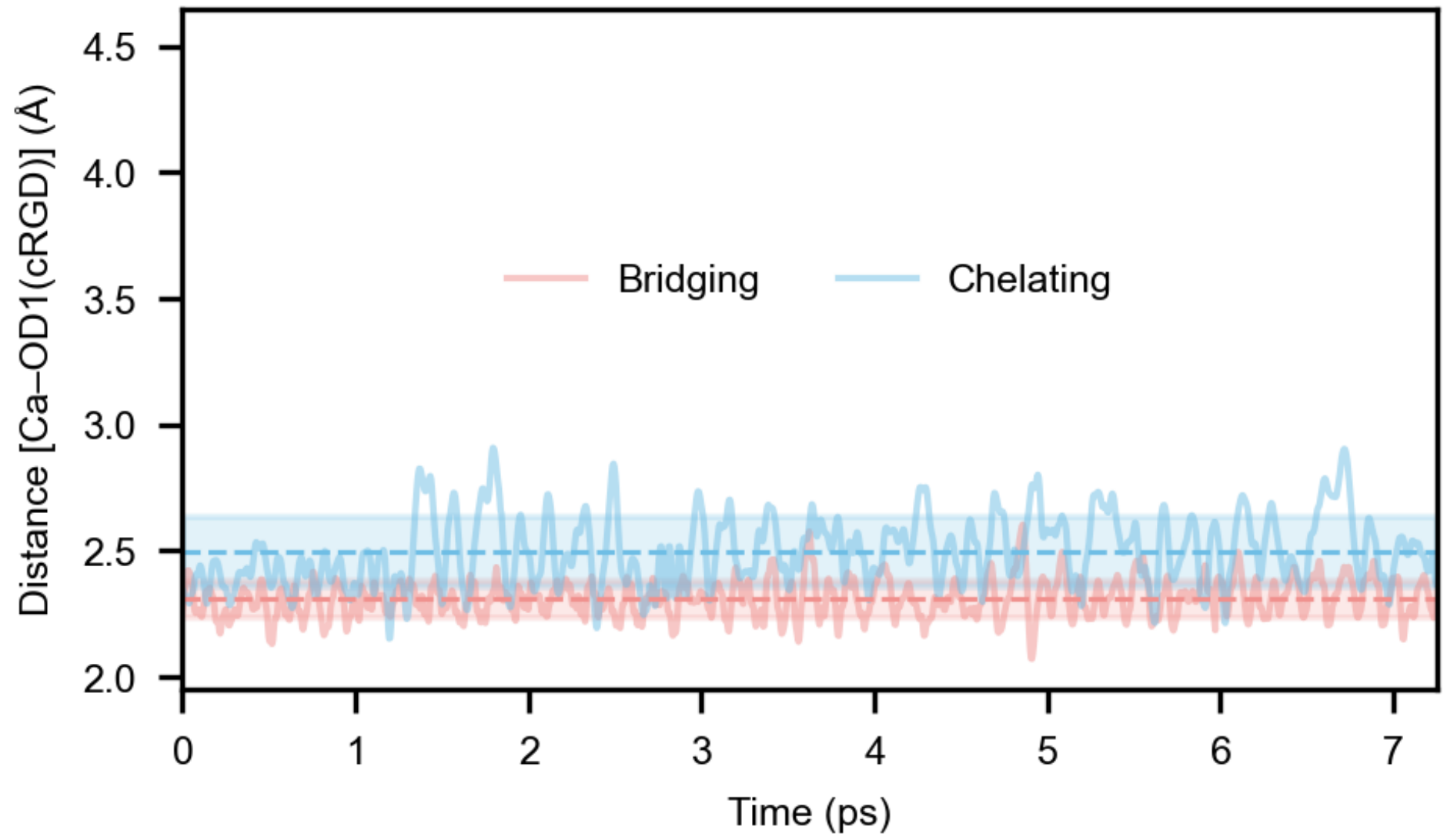

**Figure S9.** MIDAS-OD1 distance during the unconstrained QM/MM MD for the *bridging* (red) and *chelating* (blue) modes. In both cases, the average value is indicated by a dashed line, and the area of the graph within one standard deviation is highlighted with a colored bar.

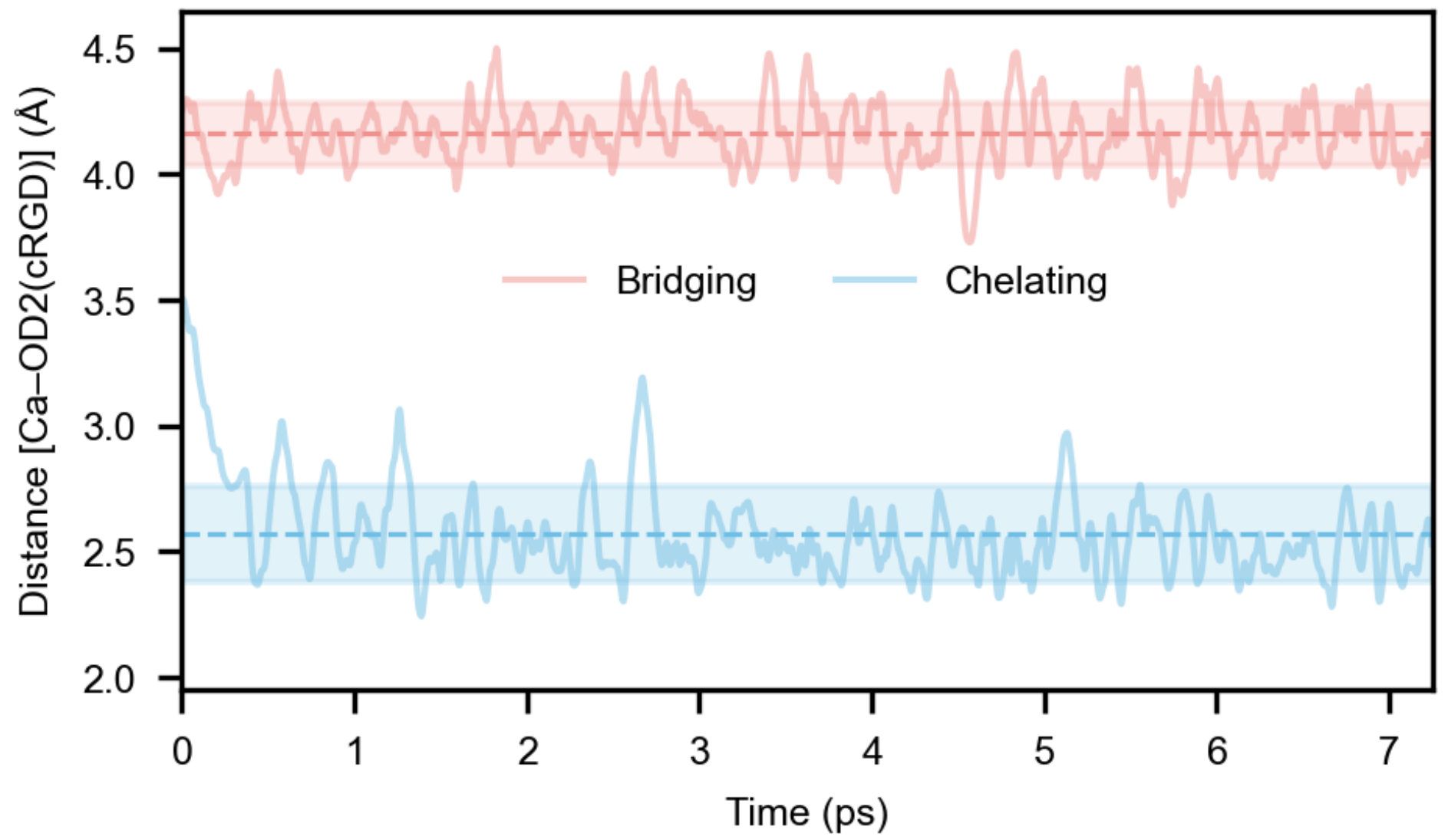

**Figure S10.** MIDAS–OD2 distance during the unconstrained QM/MM MD for the *bridging* (red) and *chelating* (blue) modes. In both cases, the average value is indicated by a dashed line, and the area of the graph within one standard deviation is highlighted with a colored bar.

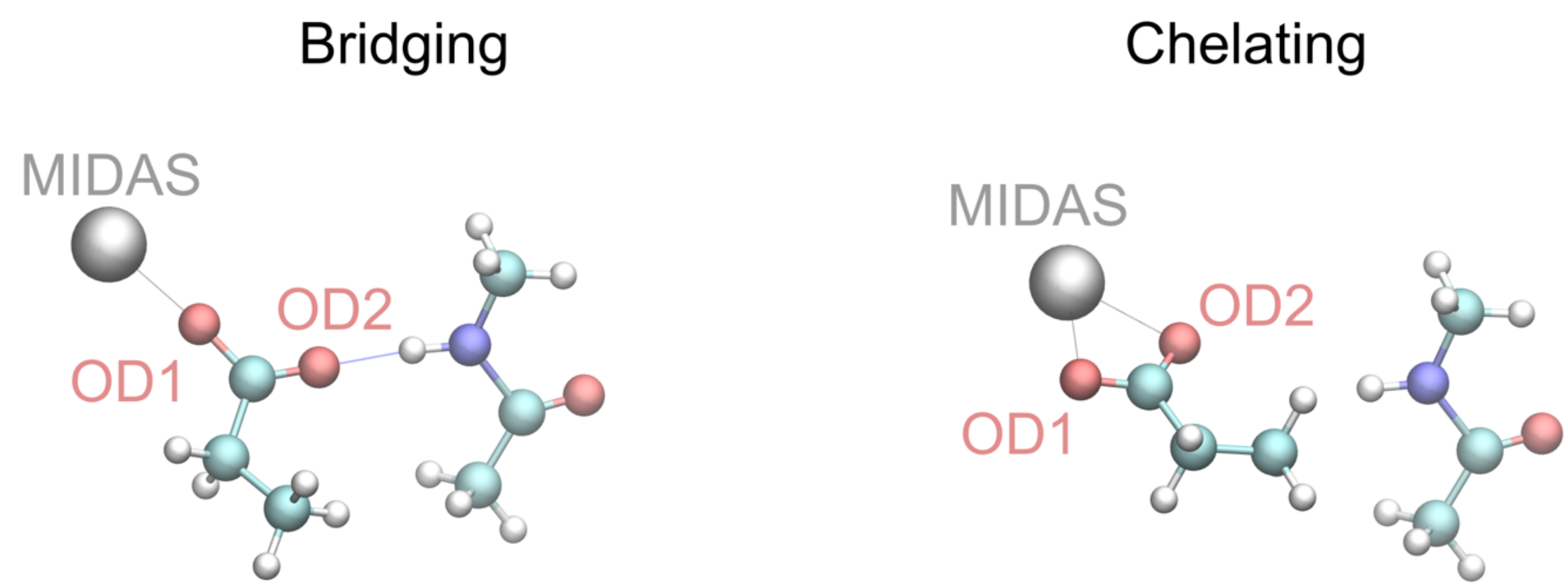

**Figure S11.** Model systems used for atomic charge analysis in the *bridging* (left) and *chelating* (right) modes. The reported conformations correspond to the optimized geometries at the MP2/6–31G(d) level. The MIDAS–OD1/2 bonds and the hydrogen bond present in the *bridging* conformation are represented in black and in blue, respectively.